\documentclass[11pt]{article}

\usepackage[final]{acl}

\usepackage{times}
\usepackage{latexsym}
\usepackage{float}

\usepackage[T1]{fontenc}
\usepackage[utf8]{inputenc}

\usepackage{microtype}

\usepackage{inconsolata}

\usepackage{graphicx}

\title{Application of LLMs to Threat Assessment\\
of Foreign Peacekeeping Missions}

\author{Gerhard Backfried, Christian Schmidt, Diego Pilutti \\
HENSOLDT Austria, Vienna, Austria \\
\texttt{\{firstname.lastname\}@hensoldt.net} \\
\AND
Michael Suker \\
Austrian Ministry of Defence, Vienna, Austria\\
\texttt{michael.suker@bmlv.gv.at} \\
}

\begin{document}
\maketitle
\begin{abstract}
We present a novel approach for applying Large Language Models (LLMs) to threat assessment in the context of foreign peacekeeping missions. Building on the PINPOINT project and its use case, the EU Monitoring Mission in Georgia, we combine an interdisciplinary risk-model with OSINT-based media collection and LLM-supported threat extraction. The proposed workflow maps media contents to mission-relevant threats, extracts structured information and applies several additional LLM-based processing steps to improve relevance and grounding. An evaluation of threats extracted from media documents shows high agreement between automatically generated results and human judgment for core aspects such as threat and mission relevance. These results indicate that LLMs provide a promising approach to support analysts in the context of peacekeeping missions.
\end{abstract}

\section{Introduction}
Within the framework of the Common Security and Defence Policy (CSDP), the European Union (EU) is presently engaged in 22 comprehensive military and civilian missions and operations. In this role, the EU contributes to international peace and security by acting collectively in crisis management, peacekeeping, and international security actions~\footnote{~EEAS Homepage -- 'Missions and operations of the EEAS',\url{https://www.eeas.europa.eu/eeas/missions-and-operations_en"}, accessed on 05/19/2026}. 

Given the intricate nature and complexity of missions, a comprehensive, thorough and meticulous risk analysis is required at the strategic, operational and tactical level. Such analyses serve as the basis for adequate preparations as well as to effectively respond to issues which may arise during mission execution.

Mission-regions pose considerable challenges, due to their diverse socio-political, demographic, linguistic, ecological, and technological situation. In these contexts, open sources emerge as a valuable source of multi-contextual information, encompassing a wide spectrum of media outlets and individuals. These sources play a crucial role in stipulating pertinent data that is essential for monitoring and comprehending the dynamic developments occurring within mission-critical areas. They typically complement other kinds of intelligence such as human intelligence (HUMINT).

\section{PINPOINT Project}
The PINPOINT project, running from February 2023 to January 2025, addressed the planning and execution of (the Austrian participation in) European missions on behalf of CSDP~\footnote{~PINPOINT -- 'National risk management for CSDP missions using Open Source Intelligence (OSINT) and position, navigation, and timing data (PNT) monitoring',\url{https://www.kiras.at/en/financed-proposals/detail/pinpoint/}, accessed on 05/19/2026}.
Six project partners from diverse backgrounds (technology, political sciences, end-users) collaborated on the development of a systematic and comprehensive risk analysis approach. This approach comprises measures and indicators based on ethical, socio-political, demographic, human rights, legal, ecological and technological information.

In tandem with the risk-model, a set of technologies from the field of OSINT as well as position, navigation, and timing data (PNT) monitoring were developed with the aim to combine their respective outputs into a joint situational picture.

Overall, the project represented an interdisciplinary endeavor that centered on the development of a methodology and model, complemented by a novel technical framework, geared towards the comprehensive collection, enrichment, and visualization of critical information for end-users to support them effectively and ensure the security of the stakeholders involved~\cite{pinpoint2024}.

%\section{Use Case EUMM}
In the initial phase of the project, several missions which Austria participated in were considered as potential use-cases. Driven by requirements of the involved end-users, the final choice was placed on the EU Monitoring Mission EUMM~\footnote{EUMM,\url{https://www.eumm.eu/}, accessed on 05/19/2026} to Georgia. This unarmed and civilian peacekeeping mission was established by the European Union following the Russo-Georgian war of 2008~\footnote{\url{https://www.diploweb.com/EUMM-Georgia-the-European-Union.html}, accessed on 05/21/2026} and remains active to this date. The mission is composed of the headquarters in Tbilisi, field offices in Mtskheta, Gori and Zugdidi, and a support element in Brussels. Only the locations in Georgia were of relevance to the work carried out within this project. 
\begin{figure*}[ht]
\centering
\includegraphics[scale=0.75]{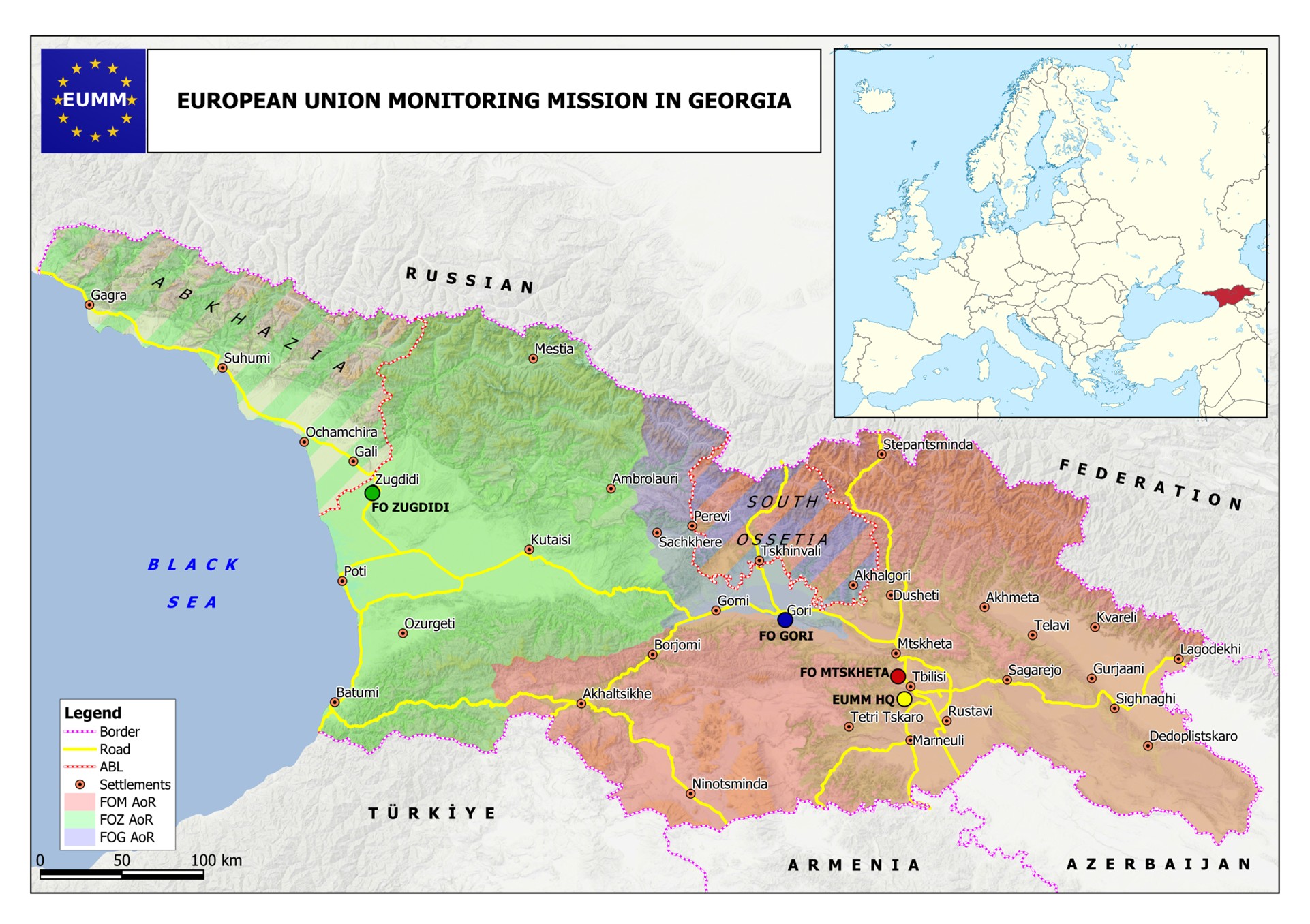}
  \caption{EUMM monitoring mission in Georgia (https://www.eumm.eu/)}
  \label{eumm_overview}
\end{figure*}
Figure~\ref{eumm_overview} presents an overview of the EUMM to Georgia.

\section{Threat Assessment} \label{threat_model}
In order to provide a comprehensive and precise representation of specific environments in which missions take place, a conceptual risk model was designed which can be used to enable an empirically grounded assessment of the conditions and influences shaping a  mission. The resulting multi-layered model is based on the analysis of a large set of factors that constitute or influence the mission-environment and draws on established approaches from the humanities and social sciences~\cite{oecd2008handbook}. 

%\subsection{Risk and Threat Model} \label{threat_model}

In order to ensure that the model is widely applicable and can be adapted to further political or geographic contexts, a clear and logical structure was devised. The modular structure breaks down the environment into a hierarchy of dimensions, categories, and indicators, enabling systematic and flexible analysis while at the same time ensuring that the model can be adapted to complex and specific requirements. 
Whereas the overall model was developed for \textit{risk assessment}, its elements can also be applied to the task of \textit{threat assessment}~\footnote{In the context of this work, \textit{threat} is a potential event or action which could cause harm or damage to the mission.}. The latter comprises the detection and analysis of threats (\textit{what could harm the mission}) whereas the former is more elaborate and also comprises the analysis of vulnerabilities and possible impacts.

The environment scanning~\footnote{The term \textit{environment scanning} was deliberately chosen (instead of \textit{environment analysis}) to emphasize the value-free nature of the process. This methodological neutrality clearly distinguishes \textit{environment scanning} from analytical evaluation methods and underscores its role as the foundation for the subsequent risk analysis.} model defines 5 dimensions as its units, which comprehensively map the mission-environment: (physical) environment, politics, society, economy, and infrastructure. These dimensions are subdivided into 26 categories, which in turn are differentiated into 151 specific indicators. Indicators represent the smallest level of analysis within the model. Threats are modeled as weighted combinations of indicators~\footnote{The model has not been published yet.}. For example, the threat \textit{Economic Dependence} is comprised of 24 indicators from 3 different categories. Indicators may originate in a variety of sources, (e.g. databases, publications, statistics) and also be \textit{inferred from media}. The latter are specifically the ones which have been considered for LLM-applicability in this work.

\section{Open Source Intelligence (OSINT)}
OSINT~\footnote{\url{https://www.natoschool.nato.int/Academics/INTEL}, accessed on 05/20/2026}, as part of an all-source strategy or complementary to other intelligence methods, such as Signal Intelligence (SIGINT) or Human Intelligence (HUMINT), has long been recognized as an efficient means of obtaining valuable and timely information~\cite{williams2018defining}. The scheme of \textit{All-Source Intelligence} comprises all available sources that can be used to gather, analyze, scrutinize and confirm publicly available information. Technological advances and the exponential increase in content production have led to the situation that some organizations in the security field already work primarily with OSINT (e.g., the UN and OSCE use OSINT in peacekeeping missions~\cite{dorn2020analysis}). 
The resulting data may be of a structured, semi-structured or unstructured nature, comes in a variety of formats, in multiple languages, dialects and modalities (text, audio, image, video, all with respective meta-data). The heterogeneity of such data requires corresponding broad capabilities for processing and analysis.
In the present work, OSINT comprises data from traditional media, such as TV, radio, or the press, as well as data from social media outlets, such as YouTube, Telegram, or VK.
The choice to use such data in the context of threat assessment is based on the hypothesis that they can act as a proxy and be employed to determine certain aspects of threat-indicators. By leveraging open sources, timely access and thorough analysis of relevant information becomes an indispensable ingredient for risk assessment, early threat detection and the creation of a comprehensive situational picture. 
However, it is also crucial to exercise judicious discretion in evaluating the credibility, accuracy, and origin of information. The clear aim of the proposed approach is thus to complement and enhance human judgment, but not to automate or replace it.

\section{Collection and Processing of Media} \label{data_collection}
Within our work, the HENSOLDT Media Mining System~\cite{backfried122015open} was employed for the collection and processing of data from a set of publicly available sources according to mission-specific requirements. Following a setup phase, contents were collected continuously and enriched with a range of technologies from the field of Natural Language Processing (NLP, LLMs) and Computer Vision (CV). The original data as well as derived information were stored and made available for further analysis and visualization. 

In order to support the selected use-case of the EUMM, a comprehensive set of sources was established and added to the Media Mining System's data-collection and processing chain. These sources reflect the geographic and contextual landscape of the setting of the EUMM and of topics relevant to the risk assessment model. To this end, more than 300 sources from Georgia, Turkey, Azerbaijan, Armenia and Russia as well as from international (governmental and non-governmental) bodies and media which report on Georgia were added. Various NLP-components and -models employed in the analysis processes were extended and adjusted to reflect the relevant topics and factors of the mission. For example, the Named Entity Recognition (NER) component was adapted to cover entities in terms of regional and contextual coverage, such as relevant locations, names of politicians and groups active in the area of interest, topics which had been issues of friction, etc. The sources and models were adapted in order to support the media-based extraction of indicators.

Next, a set of tailored queries was devised, corresponding to the specific topics and scopes of threat indicators. These queries were crafted and refined in several iterations starting with verbal descriptions of indicators and threats and guided by domain and mission knowledge. Their aim was to produce a specific set of relevant content to determine the presence or level of indicators and subsequently of a particular threat. 

Whereas the design and refinement steps took place in the initial phases of the project, the application of queries (and subsequent processing) was carried out repeatedly throughout the duration of the project. Their application yielded an up-to-date set of relevant documents forming the basis for further analysis. 
%Once triggered, the set of documents was produced in near-realtime\footnote{Near realtime means from almost instantaneously to several minutes}.
This setup allowed to determine indicator and threat levels on a continuous and timely basis.

\section{LLMs for Threat Analysis} \label{llm_for_ra}
Whereas the application of technologies from the field of AI to peacekeeping, humanitarian response and conflict mediation has received some attention over time~\cite{Olsher2015},~\cite{pasligh2019}, the application of LLMs to these areas is still in its infancy. A general consensus has been reached about the tremendous potential provided by LLMs to revolutionize information processing, decision-making and operational efficiency~\cite{caballero2025large}. However, this is met with concerns and skepticism about accountability, transparency, privacy or the tendency to completely eliminate human judgment from decision processes. While LLMs offer unprecedented analytical scale, their actual deployment requires strict safeguards and guardrails to prevent misuse and misinterpretation~\cite{johansson2025military}.
Based on the above, we embarked on the exploration of the application of LLMs to threat-analysis for peacekeeping missions. To our knowledge, this has been the first attempt to do so.

The following properties of LLMs provide strong arguments for their application in this context:
\begin{itemize}
  \itemsep0em
  \item The selected use-case is situated in a geographical area where multiple languages (and countries) are of importance. LLMs are typically able to process data in multiple languages (although not on the same performance-level) and/or can be employed for translation. 
  \item Contents in multiple formats and in different registers of language need to be processed. LLMs exhibit a degree of robustness against non-standard language, colloquialisms, typos etc. which allows them to cope with such contents.
  \item Both, the description of indicators as well as the complexity of documents require powerful semantic and reasoning capabilities to capture essential contextual information and nuances. Situations are verbally expressed in a multitude of ways; LLMs provide the ability to capture the essence and nuances.
\end{itemize}

%A fully on-premise implementation was preferred due to the sensitive nature of data as well as due to required processing capabilities in terms of volume and throughput.

Based on the model described in section \ref{threat_model}, a set of four exemplary threats was selected for investigation of the applicability of LLMs. 

% maybe need to provide a bit of info about these?
\begin{itemize}
  \itemsep0em
  \item Natural disasters: reports or media coverage reporting natural disasters or environmental incidents that could disrupt the mission's operation. On a large-scale, such incidents could also be retrieved from specific sources such as the USGS\footnote{\url{https://earthquake.usgs.gov/}, accessed on 2026/05/21}; however, for the purpose of this project and to capture local events, media-based reporting was used.
  \item External conflict actors: mentions and expressions of sentiment, hate and polarization, of external militias or groups, and of foreign actors including governments who are involved in local conflict and separatist activities
  \item Ethnic conflicts: expressions of negative sentiment, outrage or complaints against ethnic groups, individuals and local organizations
  \item Economic dependence: reporting about economic influence, manipulation, dependence or commercial activities
\end{itemize}

These threats only represent a subset of the ones whose state can potentially be determined via analysis of media contents. 
%Information leading to them can be expected to be present in the observed media. 

\section{Threat Assessment Procedure}
As outlined above, threats are associated with a set of indicators. The indicators originate from different categories. For example, the threat of \textit{conflicts with external actors} draws on more than 30 indicators from 4 categories: environment, politics, society and economy. Indicators are described verbally - the verbal description in turn forms the basis for LLM-driven analysis.

In a first step, similar to the queries which were derived from indicator descriptions, prompts were derived from the verbal descriptions of indicators associated with threats. 

The structure of the prompts was influenced by \cite{alammar2024}. These indicator-specific prompts were extended with further elements representing specific aspects of a threat (to be extracted by the LLM) such as a justification for the reported information, actors involved, locations, dates and times specifically mentioned, and an estimated severity and immediacy of a detected threat. Input documents were expected to potentially contain information about multiple threats. LLMs were applied with few-shot prompting and in reasoning mode to yield comprehensive outputs. In case of non-English documents, these were first translated into English.
%~\footnote{In a first stage, documents were processed in their original language. But as the resulting information had to be translated to English for final presentation, it was decided to do so at the start and apply all analyses to the English version. The original documents are referenced and can be consulted in both cases (should analysts speak the respective language and want to do so, e.g. for confirmation}.

Using the above approach, a set of candidate threats was produced in a structured format (JSON) in a first phase of LLM-application. The structure for each detected threat contains the following elements:

\begin{itemize}
    \itemsep0em
    \item threat: concise verbal description of the threat itself
    \item justification: verbal justification of why the threat is relevant to the mission
    \item actor: name(s) of actors involved in or causing the threat
    \item locations: names of relevant locations associated with the threat
    \item threat-level: perceived severity of the threat on a scale of 1 to 9
    \item immediacy: temporal proximity of the threat: 'short-term' when a threat is imminent within hours or a few days, 'mid-term' if a threat is likely to produce impacts within a few weeks, or 'long-term' in case a threat is likely to produce effects only within months or years or over an extended period of time
    \item date: relevant date or dates associated with the threat
\end{itemize}

This candidate set was passed through further stages of LLM-application to ground the detected information~\cite{zheng2023judging} and ensure it is pertinent to the mission. In this step, any irrelevant potential threat was removed. 

The overall sequence of LLM-applications represents a linear workflow and was implemented using LangChain~\footnote{LangChain, \url{https://www.langchain.com/}, accessed on 05/26/2026}.

Threats produced in the above manner can be filtered, ranked and included in the follow-on risk assessment procedure.

In subsequent processing, the (descriptions of the) detected and grounded threats were embedded using sentence-transformers~\cite{reimers2019} and clustered. In this manner, end-users could inspect groups of threats as well as visualize their properties, such as the temporal or geographic extensions
~\footnote{Such clusters of threats, the sources they eminate from and the points in time of publishing of underlying content currently forms the basis of further work in the realm of disinformation and propaganda.}.

\section{Evaluation}
The extraction of threats from media contents resembles an information-retrieval task, lending itself to evaluation in terms of accuracy or precision and recall. However, such an evaluation requires the presence of an annotated dataset to be used as ground truth. Due to the specific setting and novelty of the approach, no such standard test set exists. Consequently, we decided to produce a set of threats ourselves and have these rated by a group of domain experts~\footnote{Cyber Documentation and Research Center, National Defence Academy of the Austrian Armed Forces} in order to determine the accuracy of the generated results. Due to a lack of time, the \textit{reverse} setting - having a group of experts produce a set of threats and then compare these to an automatically generated set - could not be performed.

To this end, data from two time-periods - July to October 2023 and July to October 2024 - covering the four types of threats outlined above in section \ref{llm_for_ra} were selected as the basis for evaluation. Table \ref{table:threatset} provides details on the respective instances of detected threats in these documents.

\begin{table}[h!]
\begin{center} \small
\begin{tabular}{||l r r r||} 
 \hline
 Threat & 2023 & 2024 & Overall \\ [0.7ex] 
 \hline\hline
 Natural Disasters & 887 & 647 & 1534 \\ 
 \hline
 External Conflict Actors  & 71 & 206 & 277 \\
 \hline
 Ethnic Conflicts & 945 & 850 & 1795 \\
 \hline
 Economic Dependence & 281 & 406 & 687 \\
 \hline
 Sum & 4207 & 4133 & 8340 \\ [1ex] 
 \hline
\end{tabular}
\end{center}
\caption{Number and type of detected threats.}
\label{table:threatset}
\end{table}

The distribution of the different types of threats corresponds roughly between the two periods of time aligning with a stable (i.e. unchanged) situation in the area of the mission.

Fifty documents in English and Russian were selected randomly from the two time-periods, with half of the documents from 2023 and the other half from 2024. A total of 56 threats were detected automatically within these documents. These 56 threats were used for evaluation. 

To evaluate these threats, seven teams of three domain experts - all of them knowledgeable in the domain - were formed.
The teams were presented with 11 questions to be answered with \textit{yes}, \textit{no} or \textit{partially}. These questions all aimed to determine the degree of relevance of the detected threats and their individual aspects to the mission. Each detected threat was evaluated by 3 teams.

Examples of questions to be answered are:
\begin{itemize}
    \itemsep0em
    \item threat: is the detected threat indeed a threat?
    \item EUMM relevance: is the reported threat relevant to the operation?
    \item threat in text: can the threat be directly inferred from the text?
\end{itemize}
The complete set of questions can be found in annex \ref{annexa}.

For the evaluation of accuracy, each positive (yes) answer received one point, each "partially yes" answer received 0.5 points and no answers received 0 points\footnote{This does not correspond to the common definition of \textit{accuracy} but was deemed adequate for the setting as also partially relevant information should be brought to the end-user's attention for further inspection.}. Only one consolidated answer per team was allowed, to ensure robust and balanced answers. The sum of points received was then divided by the total number of possible points, yielding a resulting score between 0 and 1. Teams were invited to provide verbal feedback in addition to their answers for subsequent qualitative evaluation. 

\begin{figure}[h!]
\includegraphics[scale=0.75]{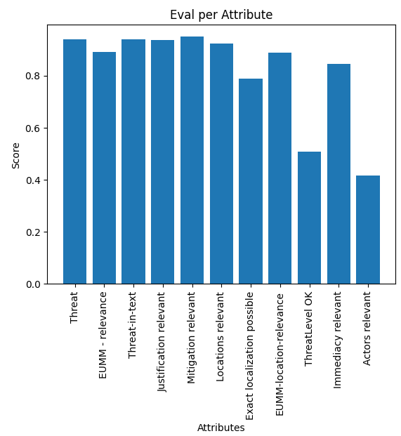}
  \caption{Evaluation Scores}
  \label{eval_scores}
\end{figure}

Overall, 48 threats were rated by the seven teams~\footnote{Unfortunately not all teams managed to rate all threats assigned to them within the time of the evaluation session. The remaining ones were not considered further.}. Figure~\ref{eval_scores} presents an overview of scores for the individual questions.

The average score reached is 0.82\footnote{0.79 if strict accuracy (not including \textit{partial} correctness) is employed.}. Several aspects obtained scores of over 0.79 indicating a high level of agreement with the automatically produced results. Questions concerning the threat itself, whether it was indeed present in the input text, relevant to the mission or whether the locations mentioned were relevant all received scores between 0.89 and 0.95 indicating the relevance and accuracy of the results produced. On the other end of the spectrum are the scores for the threat level and the actors involved. The threat level was generally perceived to be too low. The actors mentioned may be limited and/or biased due to the setting of the mission, typically groups and persons associated with the breakaway regions or Russia. 

In general, the threats themselves and their relation to the mission obtained high scores, indicating a high level of agreement between automatically generated information and human judgment. As such, these results are promising in terms of future (automated) support for further missions.

\section{Conclusion and Outlook}
We have presented a semi-automatic approach for the application of LLMs to support threat assessment in foreign peacekeeping missions. Based on an indicator-based risk model and the use-case of the EUMM in Georgia, the approach combines OSINT-based media collection, mission-specific retrieval as well as LLM-supported threat extraction and grounding. The resulting workflow transforms multilingual media contents into structured and faceted threat information.

An evaluation with domain experts yields high agreement for central aspects such as threat relevance, mission relevance and location relevance. These results indicate that LLMs can provide valuable support for analysts by improving the scalability and timeliness of risk assessment. However, lower scores for threat-level estimation and actor identification underscore the continued need for human validation and careful operational safeguards - we advocate a human-in-the-loop approach. Due to the generality of the underlying risk model and LLM-based processing, the approach lends itself to broader (and cost-effective) application to further contexts and beyond peacekeeping missions.

\section{Limitations}
Overall, we have aimed to implement a generic and robust approach, allowing to produce (media-based) indicators which can form part of a more comprehensive approach to threat assessment. However, there are several limitations to the current approach itself as well as the work carried out within the project as outlined below:

\textbf{Number and types of examined indicators}: even though the approach is based on a general model and generic in nature, it has only been investigated within a limited scope. Only four indicators have been examined for two periods of time. Whereas analysis seems to indicate a good level of agreement between the two examined time-periods, further examination as the mission evolves over should be performed. Furthermore, other indicators need be examined to support the targeted generality of the approach.

\textbf{Overall threat assessment}: using the proposed method only forms one aspect of a more general methodology for threat assessment. The combination with other elements and the interplay of the proposed method with such elements needs to be further examined. Only a combination will provide a holistic picture regarding threats.

\textbf{Mitigations}: the suggested mitigations are on a rather general level. They should take into account specifics and be tailored to the actual mission situation. 

\textbf{Robustness against manipulation or disinformation}: even though the method follows a principled approach in the selection of sources and contents as well as a qualification of sources in terms of their (perceived) level of reliability, it cannot be ruled out that manipulation and disinformation campaigns influence these sources and contents. On the one hand, automatic support to detect such manipulation, anomalies or outliers could be added. On the other hand, we emphasize the continuing need for human interaction (and supervision) in a human-in-the-loop approach.

\textbf{Single mission only}: The work undertaken here concerns only a single peacekeeping mission, with its peculiarities in terms of geography, actors and scope. Further missions may provide different requirements, priorities and insights which may lead to modifications or extensions of the proposed approach.

\textbf{Integration into practice}: whereas the approach was developed jointly with practitioners and directly discussed with members of the mission in situ, its integration into established procedures and processes needs to be examined. Mission-specific requirements and procedures would need to be taken into account. This may also entail mechanisms allowing mission staff to adjust and modify parameters without involvement of the project team (due to security or confidentiality reasons).

Overall, the implemented method seems promising as outlined during the evaluation. Follow-up activities in connection with different missions provide an alley to verify our initial findings, refine the approach and increase its robustness and field of applicability.

%\newpage
\section*{Acknowledgments}
PINPOINT was partly funded by the Austrian Security Research Programme KIRAS~\footnote{\url{https://www.kiras.at/home/}} of the Austrian Federal Ministry of Finance (BMF). The authors would like to thank all colleagues of the project as well as the staff of the EUMM in Georgia.
%Only in the final version.

\newpage
\bibliography{custom}

@book{williams2018defining,
  title={Defining second generation open source intelligence (OSINT) for the defense enterprise},
  author={Williams, Heather J and Blum, Ilana},
  year={2018},
  publisher={Rand Corporation Santa Monica}
}

@article{dorn2020analysis,
  title={Analysis for Peace The Evolving Data Tools of UN and OSCE Field Operations},
  author={Dorn, A Walter and Giardullo, Cono},
  journal={Security and Human Rights},
  volume={31},
  number={1-4},
  pages={90--101},
  year={2020},
  publisher={Brill Nijhoff}
}

@conference{backfried122015open,
  title={Open Source Intelligence for Traditional- and Social Media Sources},
  author={Backfried, Gerhard and Schmidt, Christian and Pfeiffer, Mark and Quirchmayr, Gerald and G{\"o}llner, Johannes},
  booktitle={Proceedings of the 10th International Conference on e-Business},
  year={2015}
}

@inproceedings{zheng2023judging,
  title={Judging LLM-as-a-judge with MT-bench and Chatbot Arena},
  author={Zheng, Lianmin and Chiang, Wei-Lin and Sheng, Ying and Zhuang, Siyuan and Wu, Zhanghao and Zhuang, Yonghao and Lin, Zi and Li, Zhuohan and Li, Dacheng and Xing, Eric P and others},
  booktitle={Advances in Neural Information Processing Systems},
  volume={36},
  pages={46595--46617},
  year={2023}
}

@article{Olsher2015,
  title = {New Artificial Intelligence Tools for Deep Conflict Resolution and Humanitarian Response},
  author = {Olsher, Daniel J.},
  journal = {Procedia Engineering},
  volume = {107},
  pages = {282--292},
  year = {2015},
  publisher = {Elsevier},
  doi = {10.1016/j.proeng.2015.06.083},
  url = {https://www.sciencedirect.com/science/article/pii/S187770581501036X}
}

@online{pasligh2019,
  author = {Hendrik A. Pasligh},
  title = {The Application of Artificial Intelligence for Peacekeeping},
  year = 2019,
  url = {https://thesecuritydistillery.org/all-articles/the-application-of-artificial-intelligence-for-peacekeeping},
  urldate = {2026-05-20}
}

@misc{johansson2025military,
  title         = {On the Military Applications of Large Language Models},
  author        = {Satu Johansson and Taneli Riihonen},
  year          = {2025},
  eprint        = {2511.10093},
  archivePrefix = {arXiv},
  primaryClass  = {cs.CL},
  url           = {https://arxiv.org/abs/2511.10093}
}

@article{caballero2025large,
  title={On Large Language Models in National Security Applications},
  author={Caballero, William N and Jenkins, Phillip R},
  journal={Stat},
  volume={14},
  number={2},
  pages={e70057},
  year={2025},
  publisher={Wiley Online Library},
  doi={10.1002/sta4.70057}
}

@book{alammar2024,
  title     = {Hands-On Large Language Models: Language Understanding and Generation},
  author    = {Alammar, Jay and Grootendorst, Maarten},
  year      = {2024},
  publisher = {O'Reilly Media},
  isbn      = {9781098150952},
  url       = {https://www.oreilly.com/library/view/hands-on-large-language/9781098150952/}
}

@article{reimers2019,
  author       = {Nils Reimers and
                  Iryna Gurevych},
  title        = {Sentence-BERT: Sentence Embeddings using Siamese BERT-Networks},
  journal      = {CoRR},
  volume       = {abs/1908.10084},
  year         = {2019},
  url          = {http://arxiv.org/abs/1908.10084},
  eprinttype   = {arXiv},
  eprint       = {1908.10084},
  bibsource    = {dblp computer science bibliography, https://dblp.org}
}

@inproceedings{pinpoint2024,
  author    = {Gerhard Backfried and Dorothea Thomas-Aniola and Diego Pilutti and Martin Boyer and Rüdiger Hein and Amir Tatabaei and Michael Zinkanell and Michael Suker and Philipp Agathonos},
  title     = {PINPOINT - A multidisciplinary framework for semi-automatic risk assessment in military operations and civilian missions},
  booktitle = {Proceedings of Conference on Cognitive and Computational Aspects of Situation Management 2023},
  editor    = {Kenneth Baclawski and Michael Kozak and Kirstie Bellman and Giuseppe D'Aniello and Alicia Ruvinsky and Candida Da Silva Ferreira Barreto},
  series    = {EPiC Series in Computing},
  volume    = {102},
  publisher = {EasyChair},
  bibsource = {EasyChair, https://easychair.org},
  issn      = {2398-7340},
  url       = {/publications/paper/NwLz},
  doi       = {10.29007/vnzb},
  pages     = {172-177},
  year      = {2024}}

@book{oecd2008handbook,
  author    = {{OECD} and {European Commission, Joint Research Centre}},
  title     = {Handbook on Constructing Composite Indicators: Methodology and User Guide},
  year      = {2008},
  publisher = {OECD Publishing},
  address   = {Paris},
  doi       = {10.1787/9789264043466-en},
  isbn      = {978-92-64-04345-9},
  url       = {https://doi.org}
}

\appendix
\section{Annex} \label{annexa}
Questions and attributes for LLM-based extraction.
\begin{itemize}
    \itemsep0em
    \item threat: is the reported threat indeed a threat?
    \item EUMM relevance: is the reported threat relevant to the operation?
    \item threat in text: can the threat be directly inferred from the text?
    \item justification: does the justification of why a threat is reported make sense?
    \item mitigation: are the suggested mitigation measures to counter the threat relevant?
    \item location relevance: are the reported locations relevant to the threat?
    \item location exact: is it possible to locate the reported locations precisely?
    \item location EUMM: are the reported locations relevant to the scope of the EUMM?
    \item threat level: is the reported threat level appropriate?
    \item immediacy: is the reported time-frame appropriate?
    \item actors: are the reported actors actually linked to the threat?
\end{itemize}

\end{document}